\documentclass{article}
\usepackage{graphicx} 

\usepackage[section]{placeins} 
\usepackage[left]{lineno}
\usepackage{placeins}
\usepackage{blindtext}
\usepackage{authblk}
\usepackage[left=7mm,right=15mm]{geometry} 
\usepackage{vmargin} 
\usepackage{amsmath,epsfig, amsthm}
\usepackage{url}
 \usepackage[sort,numbers]{natbib}
\usepackage{amssymb}
\usepackage{color}
\usepackage{lineno}
\usepackage{mathtools}

\usepackage{amsmath} 
\usepackage{hyperref}
\usepackage{amsmath}

\title{Bootstrap confidence intervals: A comparative simulation study}
\author{Vinícius Litvinoff Justus, Vítor Batista Rodrigues and Alex Rodrigo dos Santos Sousa}

\affil{Universidade Estadual de Campinas (UNICAMP)\\ Departament of Statistics, Brazil \thanks{Justus (viniliff@gmail.com), Rodrigues (vitor8batista11@gmail.com) and Sousa (asousa@unicamp.br)}}
\date{}

\begin{document}

\maketitle

\begin{abstract}
    Bootstrap is a widely used technique that allows estimating the properties of a given estimator, such as its bias and standard error. In this paper, we evaluate and compare five bootstrap-based methods for making confidence intervals: two of them (Normal and Studentized) based on the bootstrap estimate of the standard error; another two (Quantile and Better) based on the estimated distribution of the parameter estimator; and finally, considering  an interval constructed based on Bayesian bootstrap, relying on the notion of credible interval. The methods are compared through Monte Carlo simulations in different scenarios, including samples with autocorrelation induced by a copula model. The results are also compared with respect to the coverage rate, the median interval length and a novel indicator, proposed in this paper, combining both of them. The results show that the Studentized method has the best coverage rate, although the smallest intervals are attained by the Bayesian method. In general, all methods are appropriate and demonstrated good performance even in the scenarios violating the independence assumption.
\end{abstract}


\section{Introduction}

Bootstrap is one of the most powerful techniques to estimate properties of a given estimator, such as bias and standard error. It was originally proposed by \citet{Efron1979} in the 1970s. It is based on resampling from the empirical distribution of the data or from an assumed parametric model. Since its proposal, several applications have been described in statistics, such as regression, time series and survival analysis, among others. In statistical inference, bootstrap is also applied to perform hypothesis testing and to build confidence intervals for a population characteristic (parameter) of interest. This last application is our focus here.

There are several bootstrap-based methods to construct confidence intervals available in the literature. Some of them are based on the generation of the sampling distribution of the estimator using bootstrap, after which the confidence interval of the parameter is built by using the quantiles of this sampling distribution. Other methods adapt well known pivotal quantities such as the z and t statistics to build the interval. In these cases, bootstrap is applied to estimate standard errors. \citet{efronandtibs93} and \citet{davisonhinkley97} are good references about bootstrap (under both theoretical and practical viewpoints) and standard procedures to construct confidence intervals using bootstrap.   

A particularly interesting bootstrap variation is the so called Bayesian bootstrap, introduced by \citet{Rubin1981}. Bayesian bootstrap (or simply BB) is also based on simulating resamples generated by the data; but unlike simple bootstrap, BB can be interpreted, under the assumptions presented in \citet{Rubin1981}, as simulating directly from the posterior distribution of the parameter of interest. This is a key fact to construct confidence intervals or, in Bayesian language, credible intervals.

The paper of \citet{Rubin1981} does not address the issue of building credible intervals. However, given the posterior distribution of a parameter, it is easy to build credible intervals: any region with probability mass equal to $1 - \alpha$ is a credible region of level $1 - \alpha$, and, in particular, the quantiles $\alpha/2$ and $1 - (\alpha/2)$ can be considered the extremes of an interval. In this paper, despite the Bayesian form of generating the interval, we call them ``confidence intervalls''. Details on the interpretation of credible intervals from a Bayesian point of view can be found in \citet{Bernardo2009} and \citet{Paulino2018}, and some recent computational advances in this topic are presented in \citet{Oneil2022}.

Given the variety of proposed bootstrap-based methods to construct confidence intervals of a population characteristic, the goal of this paper is to compare and analyze the performance of some of these methods in Monte Carlo simulation studies under several scenarios of underlying distributions, sample sizes and correlation structures of the data. In this regard, several works have been published involving comparative studies of bootstrap techniques to construct confidence intervals. \citet{hall88} provided theoretical comparisons among seven confidence intervals based on bootstrap. \citet{burr94} compared the application of the bootstrap confidence intervals in the Cox model. \citet{reiser17} used boostrap confidence intervals in multi-level longitudinal data. \citet{cano18} compared several bootstrap confidence intervals with applications in hydrology. Although those studies have made meaningful contributions to the application of bootstrap to construct confidence intervals under different scenarios and data structures, the present study adds a Bayesian approach to compare against standard procedures. Moreover, we consider specific but practical scenarios of underlying distributions and also propose a simple measure to evaluate the performance of the methods based on the coverage rate and the median length of the interval.  

This work is organized as follows: Section 2 describes the considered bootstrap methods to construct confidence intervals and the scenarios of the simulation studies. Section 3 provides the results and analysis of the simulation studies. The paper finishes with final considerations in Section 4.

\section{Methodology}
The simulation studies involve five methods to obtain a $100(1-\alpha)\%$ confidence interval for a parameter $\theta$ based on bootstrap resampling, four of which are standard procedures available in the literature. The standard normal bootstrap and the Studentized bootstrap confidence intervals are based on the z and type-t statistics as pivotal quantities to build the interval. These procedures essentially require the estimation of standard errors by bootstrap. The percentile and the better bootstrap intervals (or BCa) are based on the empirical distribution of the bootstrap replicates of an estimator $\hat{\theta}$ of $\theta$. These methods are fully described in \citet{rizzo2019} and \citet{efronandtibs93}. Theoretical properties can also be found in \citet{davisonhinkley97}. Additionally, we consider a Bayesian credible interval based on bootstrap adapted from \citet{Rubin1981}. 

\subsection{Bootstrap-based methods to construct confidence intervals}
The next subsections provide brief descriptions of the considered methods.

\subsubsection{The standard normal bootstrap confidence interval}
The standard normal bootstrap confidence interval results from a simple method to build a confidence interval for the unknown parameter $\theta$, and is based on the Central Limit Theorem. Let $\hat{\theta}$ be an unbiased estimator of $\theta$ and $\mathrm{SE}(\hat{\theta})$ be its standard error. If $\hat{\theta}$ is a sample mean, then the Central Limit Theorem states that
\begin{equation}
  Z = \frac{\hat{\theta} - \theta}{\mathrm{SE}(\hat{\theta})} \xrightarrow[n \to \infty]{\mathcal{D}} N(0,1), \nonumber
\end{equation}
i.e, the random variable $Z$ converges in distribution to the standard normal distribution. Thus, an approximate confidence interval for $\theta$ is obtained by estimating the standard error $\mathrm{SE}(\hat{\theta})$ using bootstrap and applying the Z-interval. The following steps describe the method.

\begin{enumerate}
\item Calculate the estimate $\hat{\theta}$ of $\theta$ from the original data.

\item Obtain $B$ bootstrap replicates of $\hat{\theta}$, $\hat{\theta}^{(1)}, \cdots, \hat{\theta}^{(B)}$ and calculate the bootstrap estimate of the standard error of $\hat{\theta}$, denoted by $\hat{\mathrm{SE}}(\hat{\theta})$, 
$$\hat{\mathrm{SE}}(\hat{\theta}) = \sqrt{\frac{1}{B-1} \sum_{b=1}^B (\hat{\theta}^{(b)} - \bar{\hat{\theta}})^2},$$
where $\bar{\hat{\theta}}$ is the mean of the bootstrap replicates.

\item Calculate the $100(1-\alpha)\%$ Z-interval of $\theta$,
\begin{equation}\label{normal}
(\hat{\theta} - z_{\alpha/2}\hat{\mathrm{SE}}(\hat{\theta}) ; \hat{\theta} + z_{\alpha/2}\hat{\mathrm{SE}}(\hat{\theta})),
\end{equation}
where $z_{\alpha/2} = \Phi^{-1}(1-\alpha/2)$, $\Phi(\cdot)$ is the standard normal cumulative distribution function. 
\end{enumerate}

\subsubsection{The Studentized bootstrap confidence interval}
The approximation of the $Z$ statistic to build the standard normal bootstrap confidence interval is not exactly normally distributed, even if the estimator $\hat{\theta}$ is unbiased for $\theta$ and normally distributed, due the use of the bootstrap estimator $\hat{\mathrm{SE}}(\hat{\theta})$ of the standard error $\mathrm{SE}(\hat{\theta})$ in \eqref{normal}. Moreover, the t-distribution cannot be assumed to the approximated Z statistics because the distribution of the bootstrap estimator of the standard error is unknown. In this sense, the Studentized bootstrap confidence interval involves the use of a statistic based on the t statistic and its sampling distribution, which is obtained by resampling to build the $100(1-\alpha)\%$ confidence interval of the parameter $\theta$. The following steps summarize the method:  

\begin{enumerate}
    \item Calculate the estimate $\hat{\theta}$ of $\theta$ from the original data.
    \item For each replicate $b$, $b = 1,\cdots,B$:
    \begin{enumerate}
        \item Obtain the bootstrap replicate $\hat{\theta}^{(b)}$ of $\hat{\theta}$.
        \item Estimate the standard error $\mathrm{SE}(\hat{\theta}^{(b)})$ of $\hat{\theta}^{(b)}$ by bootstrapping the current bootstrap sample and calculating  $\hat{\mathrm{SE}}(\hat{\theta}^{(b)})$.
        \item Calculate the $b$-th replicate of the type-t statistic
        \begin{equation}
  t^{(b)} = \frac{\hat{\theta}^{(b)} - \hat{\theta}}{\hat{\mathrm{SE}}(\hat{\theta}^{(b)})}.  \nonumber
\end{equation}
    \end{enumerate}
    \item Calculate $\hat{\mathrm{SE}}(\hat{\theta})$ based on the bootstrap replicates $\hat{\theta}^{(1)}, \cdots, \hat{\theta}^{(B)}$.
    \item The $100(1-\alpha)\%$ confidence interval of the parameter $\theta$ is then given by
    \begin{equation}\label{t}
        (\hat{\theta} - t_{1-\alpha/2}^{*}\hat{\mathrm{SE}}(\hat{\theta}) ; \hat{\theta} - t_{\alpha/2}^{*}\hat{\mathrm{SE}}(\hat{\theta}) ),
    \end{equation}
    where $t_{1-\alpha/2}^{*}$ and $t_{\alpha/2}^{*}$ are the $1-\alpha/2$ and $\alpha/2$ sample quantiles of $t^{(1)},\cdots,t^{(B)}$ respectively.
\end{enumerate}

Note that although this procedure does not assume the normal or t distribution of the pivotal statistic, it is computationally more expensive than the other procedures since it is bootstrap nested inside a bootstrap resampling.  

\subsubsection{The percentile bootstrap interval}
The percentile bootstrap interval is obtained by generating $B$ bootstrap replicates of the estimator $\hat{\theta}$ of the parameter $\theta$ and taking the suitable quantiles of the replicates as the $100(1-\alpha)\%$ confidence interval of $\theta$. Thus, this method considers the empirical distribution of the bootstrap replicates as the reference distribution, and assumed that the empirical quantiles are estimators of the quantiles of the sampling distribution of $\hat{\theta}$. The method consists of the following steps.

\begin{enumerate}
\item Generate $B$ bootstrap replicates of $\hat{\theta}$, $\hat{\theta}^{(1)}, \cdots, \hat{\theta}^{(B)}$.

\item If $\hat{\theta}_{\alpha/2}$ and $\hat{\theta}_{1-\alpha/2}$ are the $\alpha/2$ and $1-\alpha/2$ quantiles of the generated bootstrap replicates respectively, then the $100(1-\alpha)\%$ confidence interval of $\theta$ is
\begin{equation}\label{perc}
(\hat{\theta}_{\alpha/2} ; \hat{\theta}_{1-\alpha/2}).
\end{equation}

\end{enumerate}

\subsubsection{The better bootstrap confidence interval (BCa)}
Modifications to the percentile bootstrap confidence interval method were proposed in order to improve the computational performance and to have better theoretical properties. The BCa (bias corrected and adjusted for acceleration) interval, also known as the better bootstrap confidence interval, applies adjustments in the confidence interval based on corrections for bias and skewness. It can be described by the following steps:
\begin{enumerate}
    \item Calculate the estimate $\hat{\theta}$ of $\theta$ from the original data.
    \item Generate $B$ bootstrap replicates of of $\hat{\theta}$, $\hat{\theta}^{(1)}, \cdots, \hat{\theta}^{(B)}$.
    \item Calculate the bias correction estimate $\hat{z}_0$ given by
\begin{equation}
    \hat{z}_0 = \Phi^{-1}\left(\frac{1}{B} \sum_{b=1}^{B} \mathrm{I}(\hat{\theta}^{(b)} < \hat{\theta})\right), \nonumber
\end{equation}
   where $\mathrm{I(\cdot)}$ is the indicator function.

   \item Calculate the skewness correction estimate $\hat{a}$ from jackknife replicates, given by
\begin{equation}
    \hat{a} = \frac{\sum_{i=1}^{n} (\bar{\hat{\theta}}_{(\cdot)} - \hat{\theta}_{(i)})^{3})}{6[\sum_{i=1}^{n} (\bar{\hat{\theta}}_{(\cdot)} - \hat{\theta}_{(i)})^{2})]^{3/2}}, \nonumber
\end{equation}
where $\hat{\theta}_{(i)}$ is the estimate of $\theta$ without the $i$-th observation and $\bar{\hat{\theta}}_{(\cdot)} = (1/n)\sum_{i=1}^n \hat{\theta}_{(i)}$.

\item Calculate the quantities $\alpha_1$ and $\alpha_2$ given by
\begin{equation}
   \alpha_1 = \Phi \left( \hat{z}_0 + \frac{\hat{z}_0 + z_{\alpha/2}}{1 - \hat{a}(\hat{z}_0 + z_{\alpha/2})} \right), \nonumber
\end{equation}
and
\begin{equation}
   \alpha_2 = \Phi \left( \hat{z}_0 + \frac{\hat{z}_0 + z_{1-\alpha/2}}{1 - \hat{a}(\hat{z}_0 + z_{1-\alpha/2})} \right), \nonumber
\end{equation}

where $z_{\alpha/2} = \Phi^{-1}(\alpha/2)$ and $z_{1-\alpha/2} = \Phi^{-1}(1-\alpha/2)$.

\item If $\hat{\theta}_{\alpha_1}$ and $\hat{\theta}_{\alpha_2}$ are the $\alpha_1$ and $\alpha_2$ quantiles of the generated bootstrap replicates respectively, then the $100(1-\alpha)\%$ confidence interval of $\theta$ is
\begin{equation}\label{bc}
(\hat{\theta}_{\alpha_1} ; \hat{\theta}_{\alpha_2}).
\end{equation}
\end{enumerate}

\subsubsection{Generating Bayesian bootstrap intervals}

Similarly to the simple bootstrap, the Bayesian bootstrap (BB) works under the assumption that the observed sample comes from identically distributed random variables, conditionally independent and identically distributed (iid) given a set of parameters $\beta_1, ..., \beta_n$ (for a Bayesian interpretation of this conditional independence condition, see \citet{Fortini2014} and \citet{Oneil2009}). For each observed value $x_i$, $\beta_i$ is interpreted as the probability of the value $x_i$ (and, in this sense, the BB works under the assumption that unobserved values have zero probability, see \citet{Rubin1981}).

The prior distribution of $\theta = (\beta_1, ..., \beta_n)$ is assumed to be an improper distribution, generally called Haldane's prior in the bivariate case (see, for instance, \citet{Geisser1984}), which is considered a non-informative prior. This leads to a posterior distribution being a Dirichlet distribution with the concentration parameters equal to 1, see \citet{Rubin1981}. Thus, once the parameters are simulated from the posterior distribution, the following transformations leads to the mean and standard deviation induced by the parameters:

$$\mu = \sum_{i=1}^{n} \beta_i x_i;$$

$$\sigma = \sqrt{\sum_{i=1}^{n} (x_i - \mu)^2 \beta_i}.$$

In fact, \citet{Rubin1981} also considered the case of the posterior distribution of the mean and deduced the same expression. This is an interesting fact about BB: using it, obviates the need to simulate directly from the observations $x_1, ..., x_n$; the BB provides a method to directly simulate from the posterior distribution, as proposed by \citet{Rubin1981}. In summary, the simulation of the confidence interval is done as follows:

\begin{enumerate}
    \item For $i \in \{1,2,...,B\}$, simulate $(\beta_1, ..., \beta_n) \sim Dir(n, 1, 1, ..., 1)$ (i.e, as a Dirichlet distribution with size $n$ and concentration parameters equal to 1) and set $\hat{\theta}^{(i)} = g(\beta_1, ..., \beta_n)$, where $g$ is the parameter induced by the parameters of probabilities;
    \item If $\hat{\theta}_{\alpha/2}$ and $\hat{\theta}_{1-\alpha/2}$ are the $\alpha/2$ and $1-\alpha/2$ quantiles of the generated bootstrap replicates respectively, then the $100(1-\alpha)\%$ confidence interval of $\theta$ is
\begin{equation}
(\hat{\theta}_{\alpha/2} ; \hat{\theta}_{1-\alpha/2}).
\end{equation}
\end{enumerate}

Discussion of the method can be found in \citet{Rubin1981}.

\subsection{Introducing dependence in the models through copulas}\label{copulas}

In order to evaluate the behavior of the studied methods when the independence assumption does not hold, we also generated variables with a dependence structure. This dependence was generated by copula functions, because this method allows generating dependent random variables with prescribed marginal distributions. In summary, a copula is a function that, when applied in marginal cumulative distribution functions, returns a multivariate distribution function with the prescribed marginals; for more details, see Sklar's theorem in \citet{Sklar1959} and \citet{Nelsen2006}.

It should be mentioned that a copula is always a cumulative distribution function with a standard uniform marginal distribution. This is a key fact for simulation purposes because the uniform distribution on $(0,1)$ plays an important role in simulating random variables. In this sense, once the uniform random variables $U_1, ..., U_n$ are simulated through the copula, it is sufficient to apply the transformation $F^{-1}$ in each variable, where $F$ is a target cumulative distribution function, to obtain a sequence of variables with marginal distribution $F$.

The dependence structures considered in this paper are Markovian (see \citet{Darsow1992} for a study of copulas with Markovian properties). That is, given $U_i$, the random variable $U_{i+1}$ is conditionally independent of the random vector $(U_1, ..., U_{i-1})$. The choice of this structure is based in two facts: first, this model is easy to simulate, because it is only necessary to generate $U_1 \sim U(0,1)$ to simulate from the conditionals $U_i | U_{i-1}$; and second, it is possible to use the ergodic theorem of Markov chains.

According to \citet{Paulino2018}, given an ergodic Markov chain and assuming that $g(U)$ has finite expectation, where $U$ is a variable with distribution $\pi$ and $\pi$ is the stationary distribution of the process, then

$$\frac{1}{n} \sum_{i=1}^{n} g(U_i) \xrightarrow[n \to \infty]{a.s.} E_{\pi}(g(U)).$$

Particularly taking $g(U) = I(\{ U \le x \})$, where $I$ is the indicator function, this result guarantees that the empirical distribution at a given point $x$ converges to the cumulative distribution function at that point. In the iid case, an analogous result is given by the strong law of large numbers and by the Glivenko-Cantelli theorem. This property is important to justify the choice of copula considered here, because we are studding a case where, regardless of the dependence structure, the empirical distribution converges (almost surely) to a target distribution. 

With these tools in hand, it is only necessary to specify the conditional distributions $U_i | U_{i-1}$ (or, equivalently, specify the copula). First, we use a copula that can be considered a disturbance of the independence condition: the Farlie-Gumbel-Morgenstern model, given by the expression $C(u,v) = uv + \delta uv(1-u)(1-v)$, with parameter $\delta = 1$. This family is well known for having a limited range of the dependence coefficients - for example, its Spearman coefficient belongs to the [-1/3, 1/3] interval and its Kendall coefficient belongs to the [-2/9, 2/9] range (\citet{Nelsen2006}). In fact, this model belongs to a more general family studied in \citet{Lallena2004} that represents a disturbance of the independence, and also belongs to the family of mixture of d-parameter exponential functions studied in \citet{Fernández2015}. \citet{Garcia2016} demonstrated that the family introduced in \citet{Lallena2004} maximizes the Tsallis–Havrda–Chavát entropy, which is an important case from a theoretical point of view.

Furthermore, considering a weak independence case is important because it is reasonable to expect that small deviations from independence imply small effects in the confidence intervals' behavior. Evaluating this is precisely our goal here. Therefore, the joint distribution of $(U_i, U_{i+1})$ is given by the Farlie-Gumbel-Morgenstern model, and, due to the conditional independence, a result of \citet{vv2022} guarantees that the remaining bivariate distributions are also from that family. For example, if $(U_i, U_{i+1})$ has Farlie-Gumbel-Morgenstern distribution with $\delta = 1$, then $(U_i, U_{i+2})$ has Farlie-Gumbel-Morgenstern distribution with $\delta = 1/3$, because $U_{i}$ and $U_{i+2}$ are conditionally independent given $U_{i+1}$. It should be mentioned that a smaller value of the parameter $\delta$ implies a Spearman's coefficient closer to zero, see \citet{Nelsen2006}. In other words, the dependence between $U_{i}$ and $U_{i+2}$ are weaker (in the sense of those coefficients) compared to the dependence between $U_{i}$ and $U_{i+1}$.

The simulation method can be summarized as follows:

\begin{enumerate}
    \item Simulate $U_1 \sim U(0,1);$
    \item For $i \in \{2,...,n\}$, simulate $F_i \sim U(0,1)$ (independent of previous observations) and define:
    
    $$U_i = \frac{(1 + \delta (1-(2 U_{i-1})) - \sqrt{((1 + \delta (1-(2 U_{i-1})))^2 - 4 F_i \delta (1-(2 U_{i-1}))))}}{2 \delta (1-(2 U_{i-1}))}.$$

    \item Define $X_i = F^{-1}(U_i),$ where $F$ is the target cumulative distribution function.
    
\end{enumerate}

The general expression for $U_i$ can be developed by calculating the conditional distribution of $U_i|U_{i-1}$ through a partial derivative of the copula (see \citet{vv2022}) and applying the inversion method to simulate from that. An explanation of the importance of the Farlie-Gumbel-Morgenstern model from the standpoint of its computational complexity can be found in \citet{Sriboonchitta2018}.

In addition to this, we also consider the case with $\delta = -1$. Unlike the first case, the model with $\delta = -1$ induces a negative association (in the sense of Spearman and Kendall coefficients) between two subsequent variables. Furthermore, regardless of the choice of the marginal distribution $F$, the values of those coefficients remain the same because they depend on the joint distribution only though the copula, see \citet{Nelsen2006} and \citet{Fredricks2007}.

\subsection{Simulation studies}

The simulations are carried out using the statistical software R. The impact of five factors on the intervals are measured: the estimated parameter (mean or standard deviation), the value of $\delta$ used to generate the sample (-1, 0 or 1, leading to sample autocorrelations of -1/3, 0 or 1/3 respectively\footnote{These autocorrelations are the Spearman correlations between subsequent variables, as explained in more details in subsection \ref{copulas}.}), the distribution (Exponential, Normal, Laplace, Uniform or Student-t), the sample size (10, 20 and 100) and the confidence level (0.8, 0.95 and 0.99).

Such diversity of scenarios is important to guarantee that the studied methods are not affected by phenomena such as outliers or autocorrelation. The Laplace and Student-t distributions have longer-than-normal tails, so they are widely used to deal with outliers. For instance, \citet{Valeriano2024} recently used the Student-t distribution for this purpose in the context of regression analysis. Moreover, the Exponential and Uniform distribution are important to analyse scenarios with, respectively, asymmetry and bounded support. From the fact that the support of the uniform distribution depends on its parameters, this is a classic case of a family of distributions violating the regularity conditions used in classical statistical inference, see \citet{Berger2001}. For this reason, the case of the Uniform distribution received a particular attention in \citet{efronandtibs93}, who showed the limitation of bootstrapping in estimating the maximum value of the Uniform distribution.

Furthermore, considering the autocorrelation is important because it is well known that in many cases, the autocorrelation can lead to different estimates of standard errors (see, for instance, \citet{Valeriano2024}), and the normal and Studentized confidence intervals are based on estimates of the standard errors (\citet{efronandtibs93}). To evaluate this, we used copula models to introduce dependence between the variables without changing their respective marginal distributions. For a technical discussion about the limitations of the bootstrap methods, see \citet{Bickel1981}.

We chose the theoretical mean to be 5 in all cases, as well as the theoretical standard deviation. This choice was based on the fact that one of the selected distributions is the Exponential distribution, in which the mean is always equal to the standard deviation. The number of bootstrap samples for each method was set to be 100 and the random seed was chosen to be 42 to ensure reproducibility. The code used to simulate the data and to construct the confidence intervals can be obtained by the reader upon request.

For each combination of the variables related to the sample (distribution, $\delta$ value and sample size) 100 samples were generated. All methods were applied to all samples, for all three chosen confidence levels. 

\section{Results and Discussion}

The methods were compared with respect to three features: coverage rate, median interval length and an indicator that combines both coverage and length information in order to assess interval quality. For each of these features, their values were compared between the methods in the general case and for each considered factor.

\subsection{Coverage rate}

\begin{figure}[!htb]
\caption{General coverage rate by method.}
\centering
\includegraphics[width=0.6\textwidth]{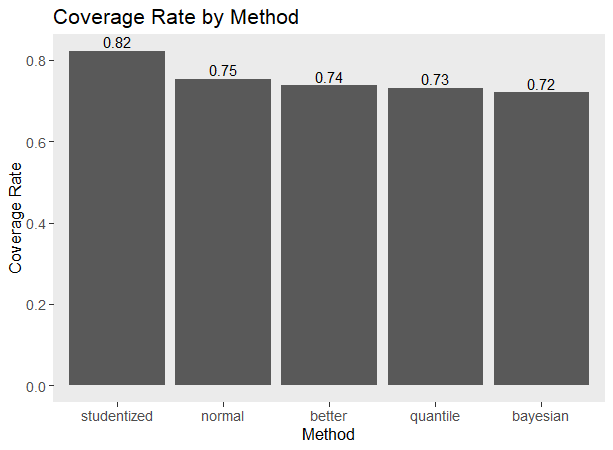}
\label{f1}
\end{figure}

Figure \ref{f1} shows the coverage rate of each method across all scenarios and its replications. It can be seen that the Studentized intervals achieved the greatest coverage rate in the big picture. The other four methods achieved fairly similar results. In the next sections of this chapter, we will show why the Studentized method achieved this big advantage.

\begin{figure}[!htb]
\caption{Coverage rate by method and under mean and standard deviation parameters.}
\centering
\includegraphics[width=0.75\textwidth]{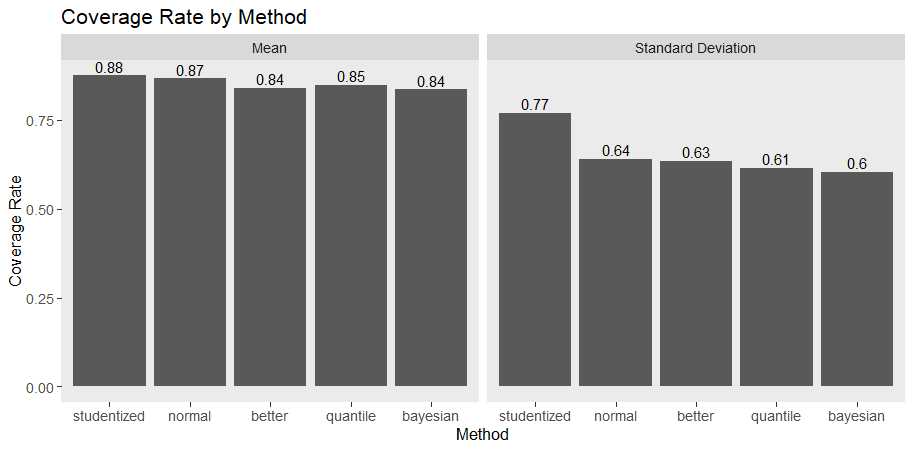}
\label{f2}
\end{figure}

Figure \ref{f2} shows that when filtering only for mean intervals, all methods reached very close results, although the Studentized method was still the best with respect to coverage rate. With regard to standard deviation intervals, however, Studentized intervals had an even greater advantage.
Also, the coverage rate is considerably smaller in the confidence intervals of the standard deviation (in comparison to the intervals of the mean) in all the methods. However, the coverage of the Studentized interval did not change as much as the other methods among the estimated parameters.

\begin{figure}[!htb]
\caption{Coverage rate by method with the values of $\delta$ used to generate the samples.}
\centering
\includegraphics[width=1\textwidth]{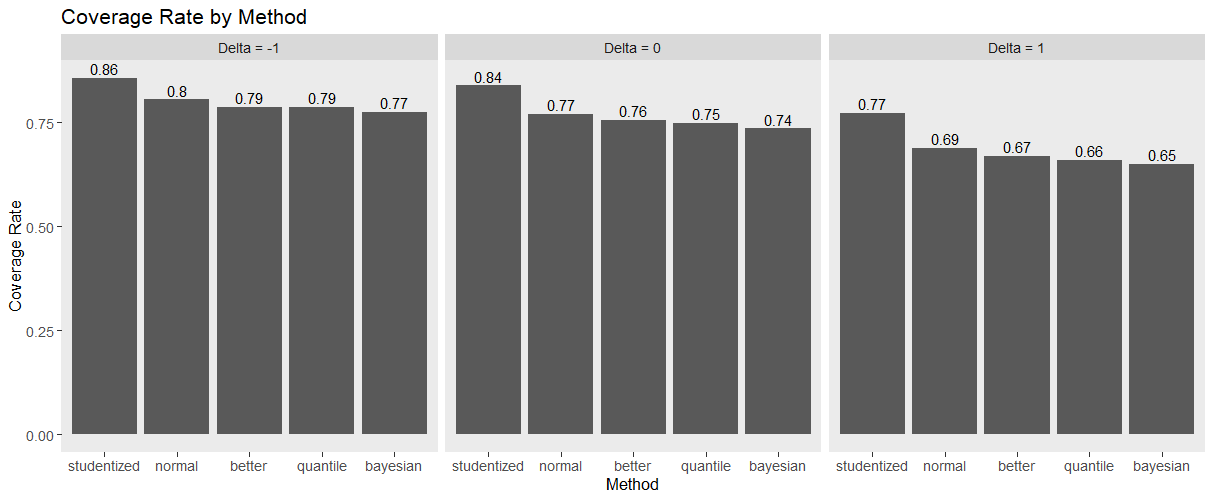}
\label{f3}
\end{figure}

Figure \ref{f3} shows that when filtering for the different chosen values of $\delta$, the order between methods did not change. The Studentized method achieved the greatest coverage rate for all chosen values of $\delta$, with the other four methods reaching close results.
However, even when the order between intervals did not change, the autocorrelation affected their behavior. The parameter $\delta$ systematically affected the coverage rates of the intervals.
Figure \ref{f3} suggests that the increase of the autocorrelation decreased the coverage rates. Indeed, this change is reasonable, given that the positive autocorrelation can decrease the dispersion of the variable, which in turn can disastrously lead to samples with very similar observations. It should be noted that all the methods were constructed under the assumption of independence. Possibly for the same reason, the negative correlation between two subsequent variables led to better coverage rates, because informally this phenomenon guarantees samples with a greater diversity of values. However, this is not an obvious result, given that the dependence model used here did not necessarily induce negative association between two non-subsequent variables.

\begin{figure}[!htb]
\caption{Coverage rate by method and sample distribution.}
\centering
\includegraphics[width=1\textwidth]{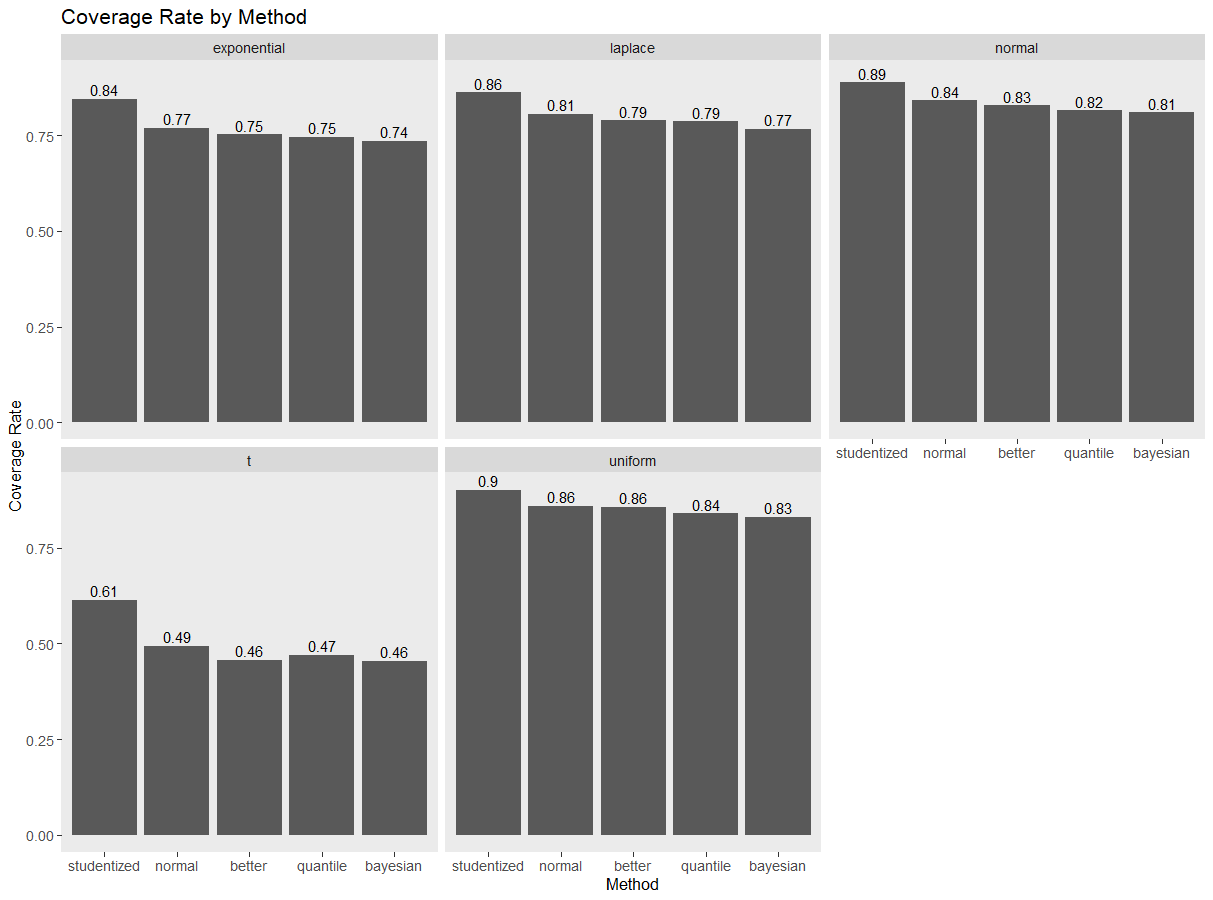}
\label{f4}
\end{figure}

When accounting for the different sample distributions, it is possible to see a similar pattern in all of them. This is shown in Figure \ref{f4}. The Studentized method reached the greatest coverage rate in all scenarios and the order of the other four methods remained approximately the same. Some distributions (such as Student-t) had a greater advantage by the Studentized method, while others (like the Uniform) showed more similar results between methods.
It is notable that the Student-t distribution had the worst confidence intervals, according to their respective coverage rates (all above 0.65). On the other hand, all the methods had at least  coverage of 0.80 for the Uniform distribution. These results are an aggregation of all scenarios with each method. This does not imply that specific scenarios cannot have coverage rates out of these bounds.

\begin{figure}[!htb]
\caption{Coverage rate by method and sample size.}
\centering
\includegraphics[width=1\textwidth]{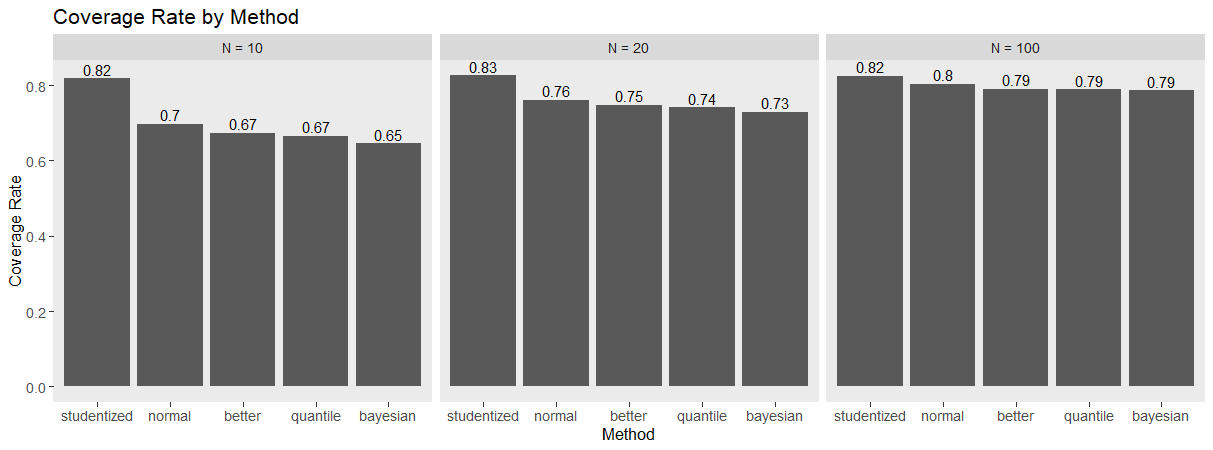}
\label{f5}
\end{figure}

Figure \ref{f5} shows an interesting finding: the Studentized method’s advantage decreased with sample size. For a sample size of 10, it reached by far the greatest coverage rate. For a sample size of 100, however, it was just three percentage points better than the worst method with respect to coverage rate, which was the Bayesian method.
Except the Studentized method, all the methods had considerably higher coverage rates. However, the coverage rate of the Studentized method was practically the same for N=10 and N=100.

\begin{figure}[!htb]
\caption{Coverage rate by method and confidence level.}
\centering
\includegraphics[width=1\textwidth]{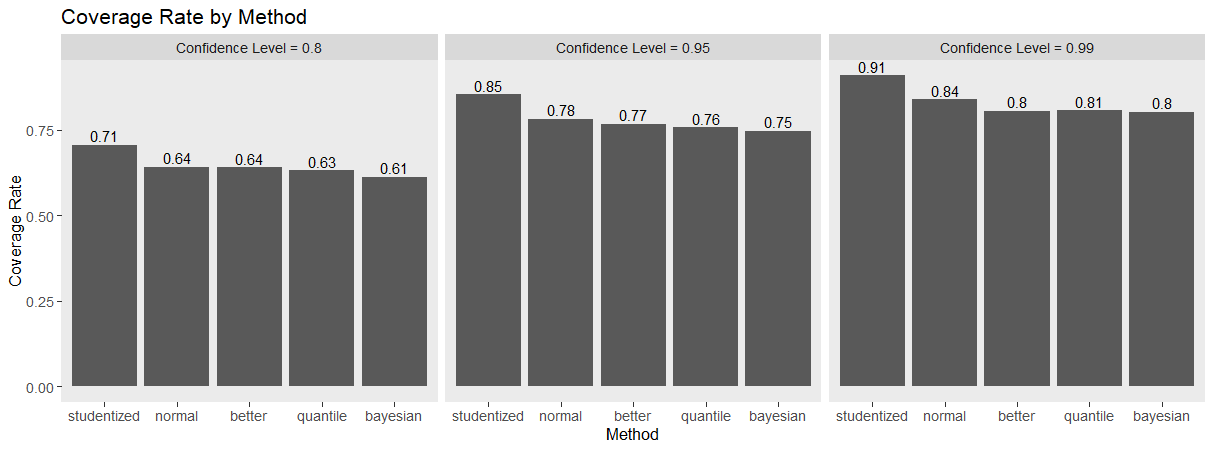}
\label{f6}
\end{figure}

Figure \ref{f6} shows that, as expected, the confidence level had no impact on the comparison between methods. The Studentized method had an advantage for all chosen levels, with the order of the other methods being approximately constant. In addition, the coverage rate increased with confidence level, as expected.

\subsection{Median interval length}

\begin{figure}[!htb]
\caption{Median interval length by method}
\centering
\includegraphics[width=0.6\textwidth]{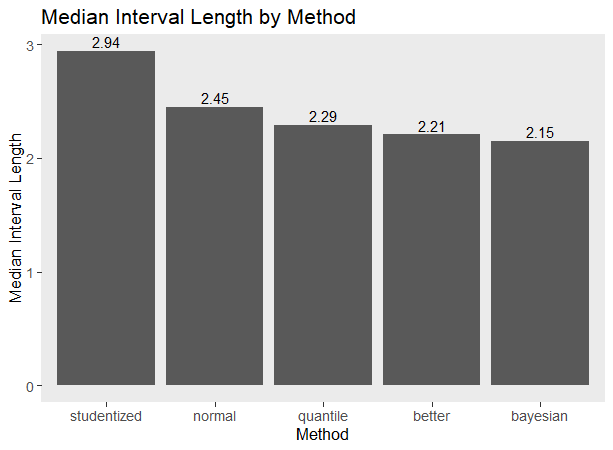}
\label{f7}
\end{figure}

Figure \ref{f7} shows that the Studentized method led by far to the longest intervals. This may explain why they stand out with respect to coverage rate. It is also important to note that the order between the methods was almost the same when comparing the coverage rates, with just the Quantile and Better methods switching places.

\begin{figure}[!htb]
\caption{Median interval length by method, when considering mean and standard deviation intervals}
\centering
\includegraphics[width=0.75\textwidth]{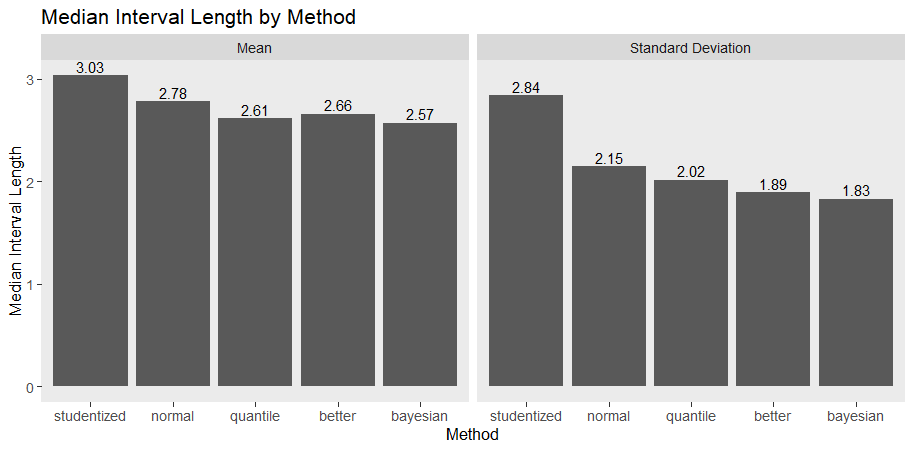}
\label{f8}
\end{figure}

When we considered the parameter for which the interval was built (Figure \ref{f8}), the Studentized method generated by far the longest intervals for both parameters. This difference was more evident, however, for standard deviation intervals. It is important to remember that the Studentized method’s advantage with respect to coverage rate was not that big for mean intervals. Hence, the method is possibly not the best in this case. This analysis will be continued further in this chapter.
We previously found that the coverage rate was better to the mean than for the standard deviation. On the other hand, Figure \ref{f8} shows that the lengths of the intervals were smaller for the standard deviation. 

\begin{figure}[!htb]
\caption{Median interval length by method, when considering the values of $\delta$ used to generate the samples}
\centering
\includegraphics[width=1\textwidth]{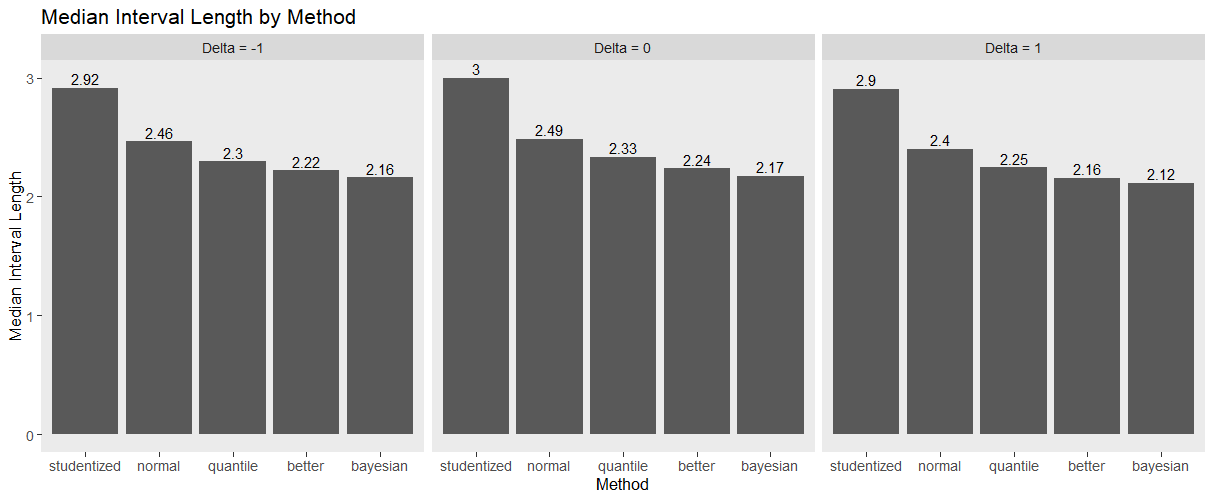}
\label{f9}
\end{figure}

Figure \ref{f9} shows that the chosen value of $\delta$ (which, as explained, was closely related to sample autocorrelation) did not have an impact on the median interval length. The three graphs are almost identical.

\begin{figure}[!htb]
\caption{Median interval length by method, when considering sample distribution}
\centering
\includegraphics[width=1\textwidth]{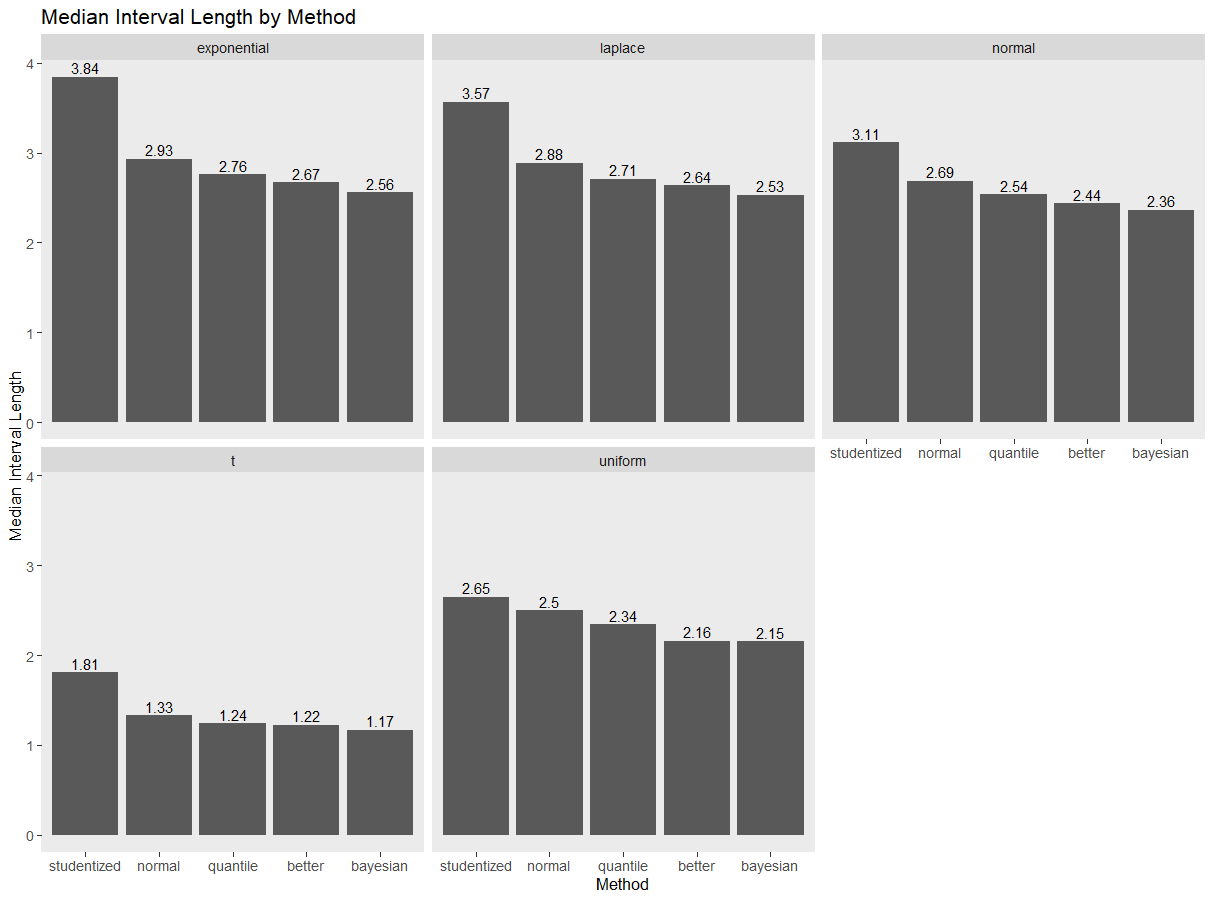}
\label{f10}
\end{figure}

In Figure \ref{f10} it is possible to notice that the order between the methods remained constant with respect to interval length across all distributions. In some distributions, however, the difference between methods was smaller, such as the Uniform distribution.

The interval lengths were considerably smaller for the Students-t distribution. However, the t distribution also achieved the worst coverage rates, so these smaller intervals cannot be considered indicators of an advantage. These facts suggest that heavy tails have a big impact on the inferential process. It is also interesting to note that, although the greatest coverage rate was observed previously in the case of the Uniform distribution, their intervals were not longer than the intervals of other distributions (except the t-distribution, which had the smallest intervals). The fact that the Uniform distribution was the only one considered in this simulation with bounded support, strengthens the assumption that the tails have an important role in the behavior of the confidence intervals.

\begin{figure}[!htb]
\caption{Median interval length by method when considering sample size}
\centering
\includegraphics[width=1\textwidth]{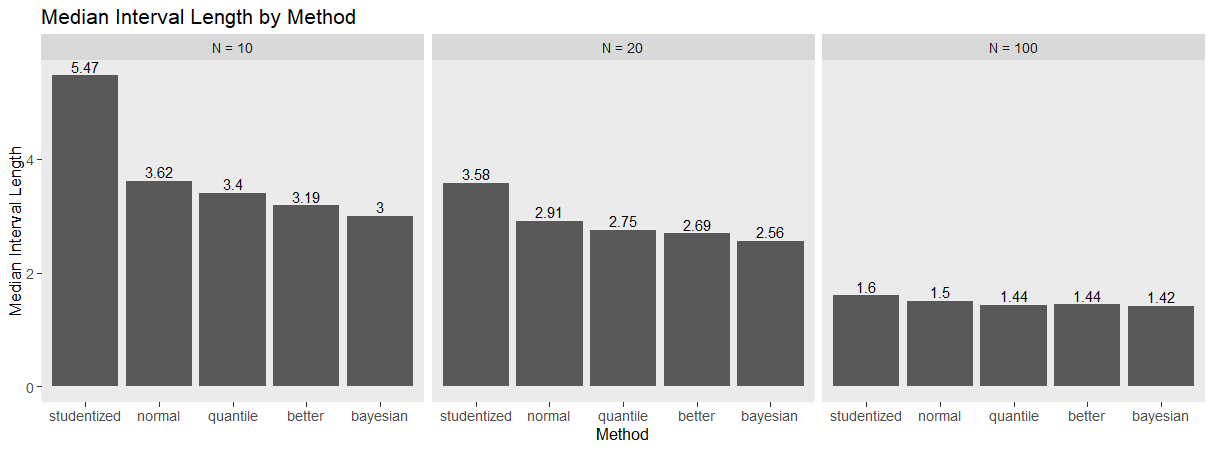}
\label{f11}
\end{figure}

The difference between methods with respect to interval length  decreased with sample size, as shown in Figure \ref{f11}. For samples of size 10, the Studentized method led to the longest intervals by far. When considering samples of size 100, nevertheless, the median interval length was very close for all methods.
For all the methods, the median interval length in the case of N=100 was less than  half the median length of the same methods for N=10.

\begin{figure}[!htb]
\caption{Median interval length by method when considering the confidence level}
\centering
\includegraphics[width=1\textwidth]{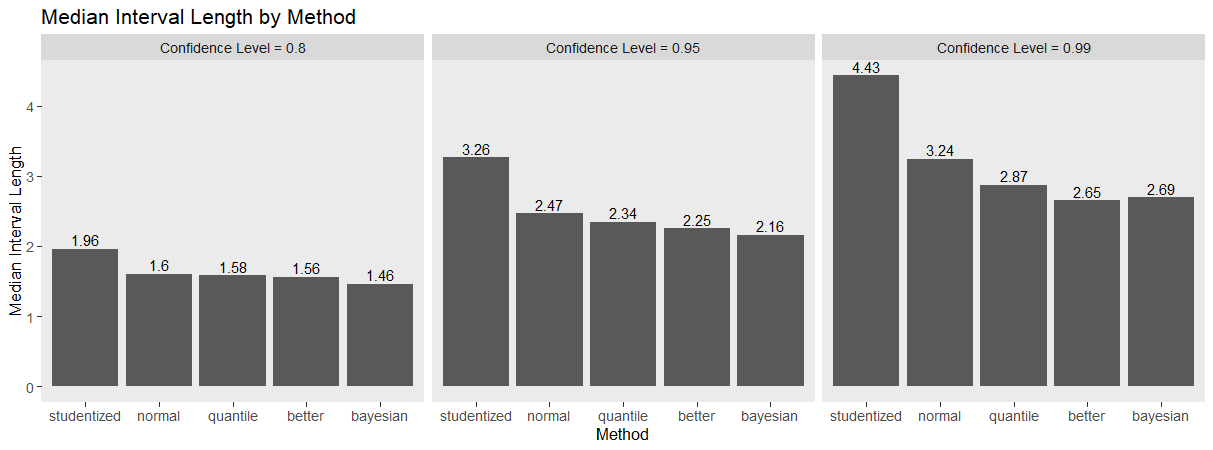}
\label{f12}
\end{figure}

Figure \ref{f12} shows that the opposite occurred with confidence level: As confidence level increased, the difference between methods also increased. Even when considering differences in percentage terms, some change occurred: for a confidence level of 0.8, the median interval length was 34.2\% greater for the Studentized intervals than for the Bayesian intervals (which are the smallest median intervals). For a confidence level of 0.99, however, this difference became 64.7\%.

\subsection{Confidence interval performance indicator}

In the previous two subsections, we showed that some methods had greater coverage rates (Studentized and normal methods) while others had smaller median interval lengths (Bayesian and Better methods). However, this did not answer the main concern of this paper: which method is the best in each case?

In order to answer this question, we created a performance indicator. This indicator takes account of both the coverage rate and the interval length as both of these variables are important. A method that reaches a huge coverage rate by generating enormous intervals has little value, as does a method that generates tiny intervals and low coverage rates. So the idea is to use a method that is balanced between these factors.

The indicator is given by the following formula: $$\frac{C}{\sqrt{1+L}},$$

where $C$ is the coverage rate and $L$ is the median length. This indicator varies between 0 and 1. It is close to 0 when either the coverage rate is low or the median interval length is high, while it is close to 1 when the coverage rate is high and the median interval length is low.

Initially, the idea was to use this indicator without the square root in the denominator. Nevertheless, while analyzing the simulated samples, we decided to check whether the indicator was giving more weight to one of the aspects, according to the Spearman correlation between the indicator and each of the aspects.

So, we grouped the data by each scenario and coverage rate and median interval length, and calculated this indicator. The estimated correlations were around -0.56 for interval length and around 0.43 for coverage rate. Hence, to balance the indicator, we inserted a square root into in the denominator. As a result, the correlations became approximately -0,43 for interval length and 0,64 for coverage rate. This new version seemed to be more appropriate for the analysis, since the coverage rate is generally more important than the interval length.

\begin{figure}[!htb]
\caption{Performance indicator by method}
\centering
\includegraphics[width=0.6\textwidth]{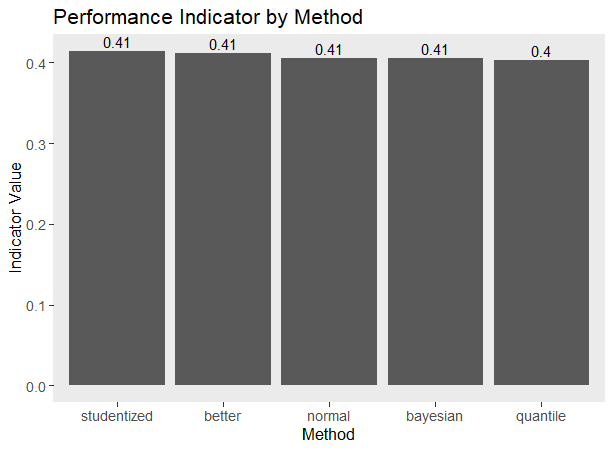}
\label{f13}
\end{figure}

Figure \ref{f13} shows that, for the general case, all methods produced very close performance indicator values. This means that the most interesting results will be found when analyzing the impact of each variable used in the simulation.

\begin{figure}[!htb]
\caption{Performance indicator by method, when considering mean and standard deviation intervals}
\centering
\includegraphics[width=0.75\textwidth]{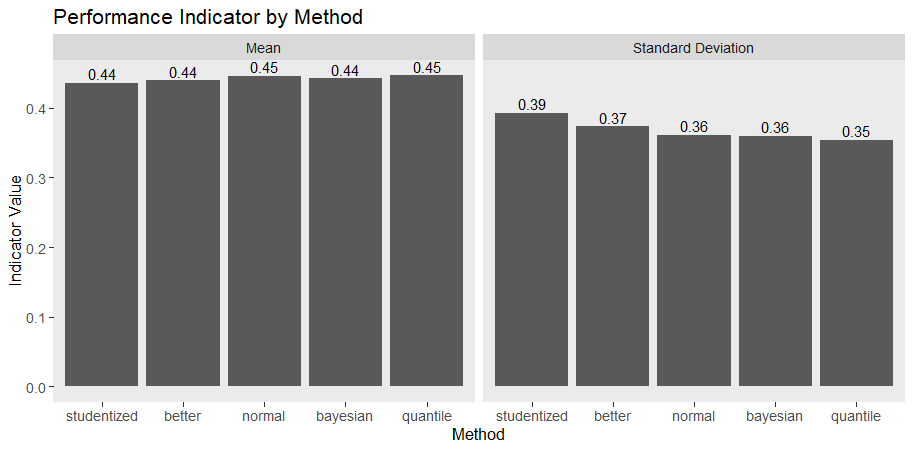}
\label{f14}
\end{figure}

Figure \ref{f14} shows that, when considering standard deviation intervals, the Studentized method had some advantage against the other methods. For mean intervals, however, all cases reached very close results, with the normal and quantile methods having slight advantages.

As previously mentioned, the intervals, in general, had better coverage and smaller length for the mean. For this reason, it is not obvious what parameters are easier to estimate by confidence intervals. The indicator suggests an advantage in the case of the mean.

\begin{figure}[!htb]
\caption{Performance Indicator by method, when considering the values of $\delta$ used to generate the samples}
\centering
\includegraphics[width=1\textwidth]{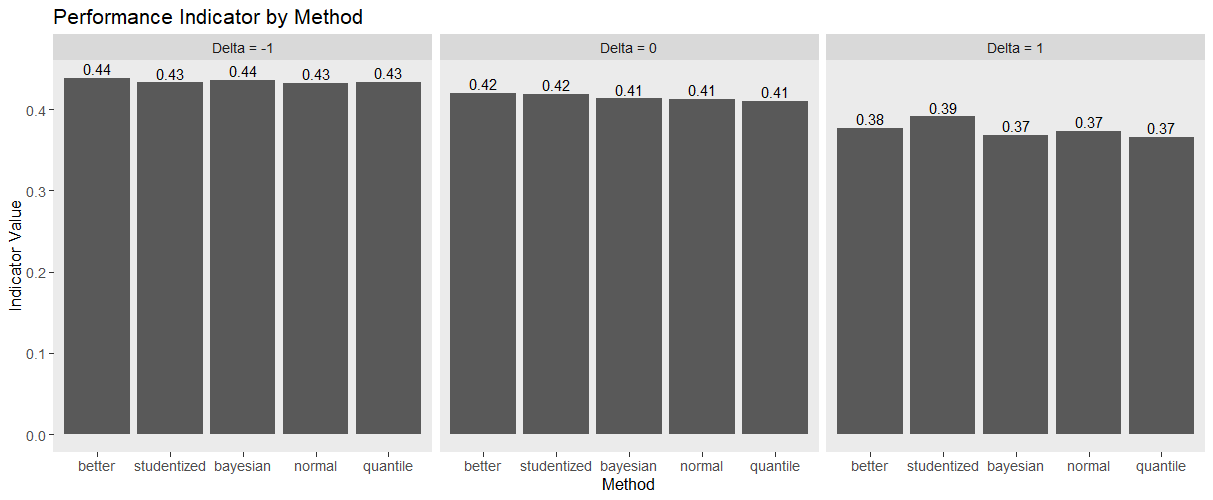}
\label{f15}
\end{figure}

In Figure \ref{f15}, it is possible to see that for $\delta$ equals -1 or $\delta$ equals 0 (independent case) all methods had very similar performance, with the Better method having slightly greater performance. For $\delta$ equals 1, however, the Studentized method showed some advantages.

The previous results showed that the interval lengths were practically not affected by the autocorrelation parameter, but the coverage rate decreased with an increase in the parameter $\delta$. Thus, as expected, the indicator was greater for lower autocorrelations.

\begin{figure}[!htb]
\caption{Performance Indicator by method, when considering sample distribution}
\centering
\includegraphics[width=1\textwidth]{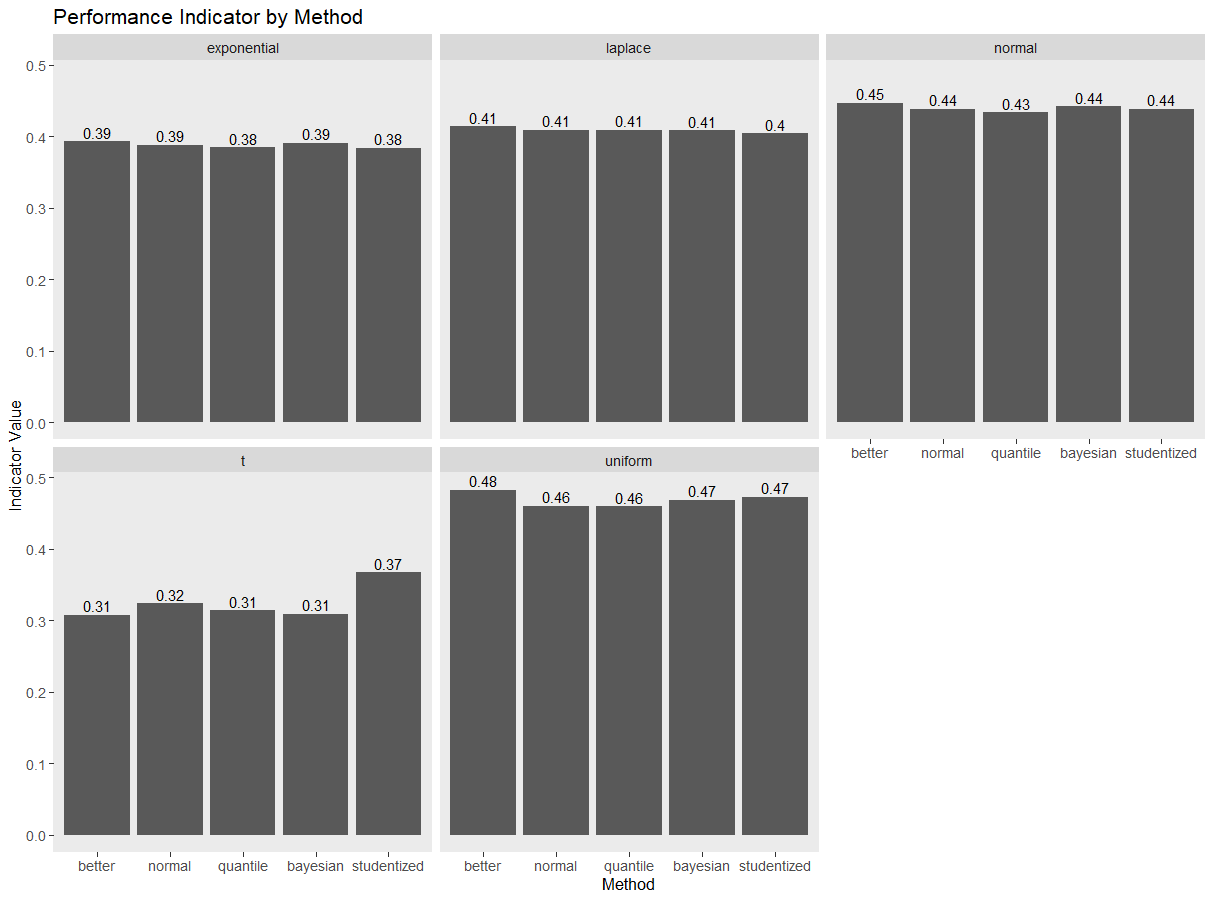}
\label{f16}
\end{figure}

When analyzing the difference between distributions in Figure \ref{f16}, it is possible to notice that the Studentized method had some advantage for Student-t distribution. For the Normal and Uniform distributions, the Better method was the best one. For Exponential and Laplace distributions, however, all methods had almost identical scores.

The indicator differs considerably in the case of the Student-t distribution. However, the advantage of the Studentized method should be viewed with caution, as it will become clear in the next subsection.

\begin{figure}[!htb]
\caption{Performance indicator by method, when considering sample size}
\centering
\includegraphics[width=1\textwidth]{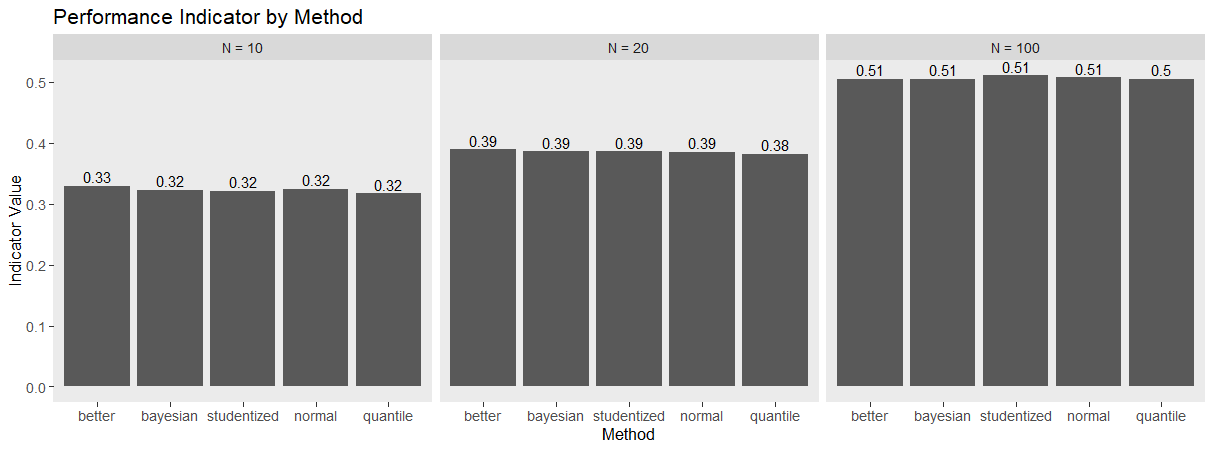}
\label{f17}
\end{figure}

Sample size did not impact the comparison between the methods with respect to the indicator, as shown in Figure \ref{f17}. All methods achieved very close results for all chosen sample sizes.
As expected, the indicator became larger with increased sample size.

\begin{figure}[!htb]
\caption{Performance indicator by method, when considering confidence level}
\centering
\includegraphics[width=1\textwidth]{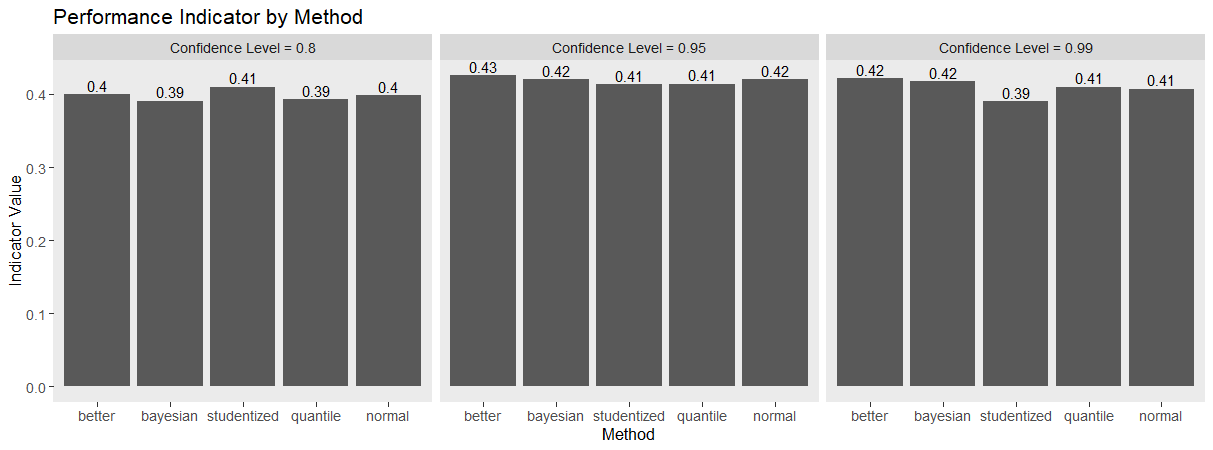}
\label{f18}
\end{figure}

Figure \ref{f18} shows an interesting finding. For a low confidence level (e.g 0.8), the Studentized method had some advantage. For the standard confidence level (0.95), the Better method beat the Studentized method. For a high confidence level, the Studentized method had the lowest score, with the Better and Bayesian methods reaching the best values.

\subsection{A note on average interval lengths and tail probabilities}

The previous subsections showed an interesting contrast: the method with the best coverage rate (Studentized) had the worst median interval length, while the method with the best median interval length (Bayesian) had the lowest coverage rate. These facts illustrate the difficulty - not restricted to the universe of confidence intervals - of comparing statistical methods. The indicator presented in the section above is a first step seeking to offer a more systematic approach to compare the methods. However, like any other approach, an indicator can hide important information. First, it should be noted that the length of a random interval is a random variable, and, in this sense, the median cannot alone capture all relevant information involving the distribution.

For this reason, in this subsection we summarize in a more informative way the distribution of the interval lengths for each method. With this information, we will show that the Studentized method systemically produced large intervals, so a person wanting to avoid such big intervals can prefer the intervals produced by the Bayesian and the Better methods.

\begin{figure}[!htb]
\caption{Boxplots of interval length by method}
\centering
\includegraphics[width=0.6\textwidth]{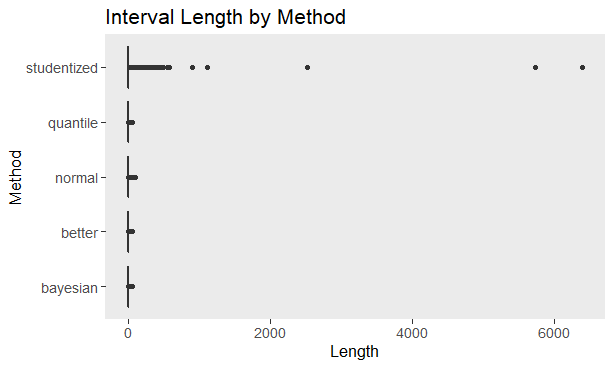}
\label{f19}
\end{figure}

Figure \ref{f19} shows boxplots of the distribution of the lengths in all the simulated intervals in all scenarios. It is noteworthy that the Studentized distribution produced extreme values, such that it is hard to recognize the presence of the boxes in the image.

\begin{figure}[!htb]
\caption{Boxplots of interval length by method, when filtering for N = 100}
\centering
\includegraphics[width=0.6\textwidth]{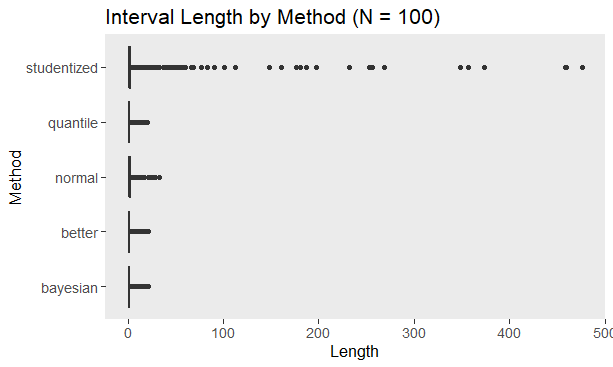}
\label{f20}
\end{figure}

The extreme values became less frequent for larger samples. However, Figure \ref{f20} shows that, even when restricting this aspect to N=100, the Studentized method produced extreme values with considerable frequency.

When filtering the intervals of the simulation with length bigger than 100, we discovered interesting findings. First, almost all of those 105 intervals were produced with the Studentized method (104) and were estimating the standard deviation (100). Second, one of these intervals was produced by the Normal method. Moreover, almost all distributions appeared in those data: the Student-t (69), the Exponential (28), the Laplace (7) and the Normal (1) distributions. Only the Uniform distribution did not have any interval with length larger than 100 in this simulation study. These facts suggest again that heavy tails induce undesirable behavior in the intervals; the Studentized method produces very wide intervals; and the standard deviation is a parameter that is harder to estimate.

\begin{figure}[!htb]
\caption{Boxplots of the scenarios mean interval length by method}
\centering
\includegraphics[width=0.6\textwidth]{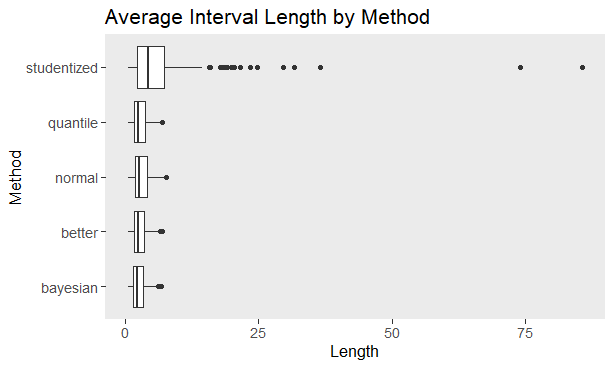}
\label{f21}
\end{figure}

Figure \ref{f21} provides a new look for the interval lengths: for each scenario, the average length was computed, and the distribution of these average lengths is displayed for each method in the figure. The Studentized method led to greater lengths.

\begin{figure}[!htb]
\caption{Boxplots of interval length by parameter}
\centering
\includegraphics[width=0.6\textwidth]{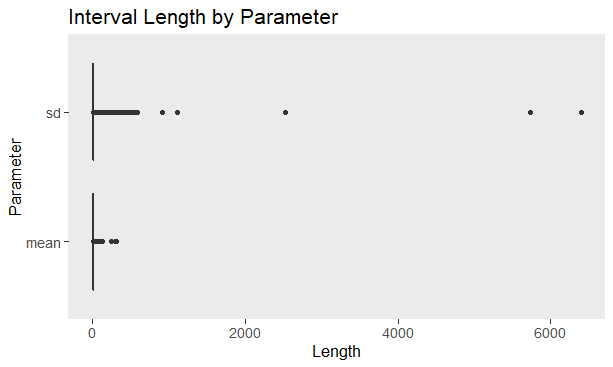}
\label{f22}
\end{figure}

\begin{figure}[!htb]
\caption{Boxplots of interval length by method, when filtering for the estimation of the mean}
\centering
\includegraphics[width=0.6\textwidth]{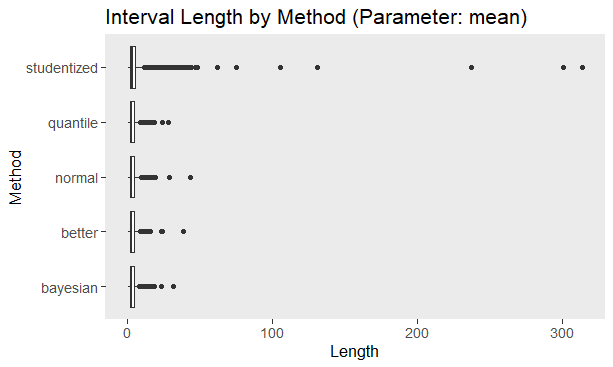}
\label{f26}
\end{figure}

Figure \ref{f22} illustrates the fact that the widest intervals were produced when the goal was to estimate the standard deviation. However, this is not necessarily an argument for the use of the Studentized method for location parameters such as the mean; even for the mean, the median length of the intervals produced by the Studentized method was bigger, as shown in a previous subsection. Figure \ref{f26} confirms that even when restricting for the mean, the extremely large intervals were more commonly produced by the Studentized method.

\begin{figure}[!htb]
\caption{Boxplots of interval length by method, when filtering for the estimation of the mean and for N = 100}
\centering
\includegraphics[width=0.6\textwidth]{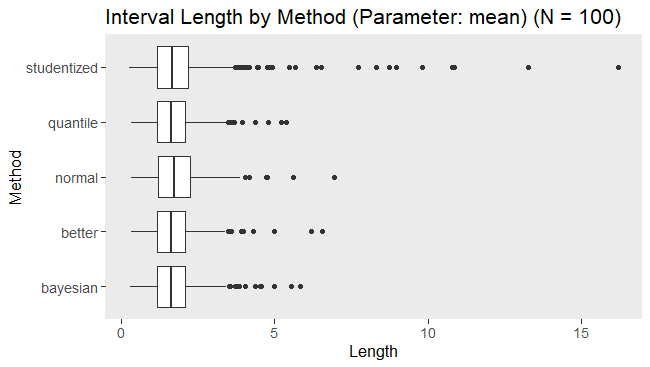}
\label{f27}
\end{figure}

Figure \ref{f27} shows that even when restricting the focus to  estimation of the mean and also to the sample size $N = 100$, the largest intervals were produced by the Studentized method. Although the largest intervals in this case were not as long as the intervals for the standard deviation, it is clear that the length of the intervals of the Studentized method had tail probabilities very different from those of the other methods.

\begin{figure}[!htb]
\caption{Mean interval length by method}
\centering
\includegraphics[width=0.6\textwidth]{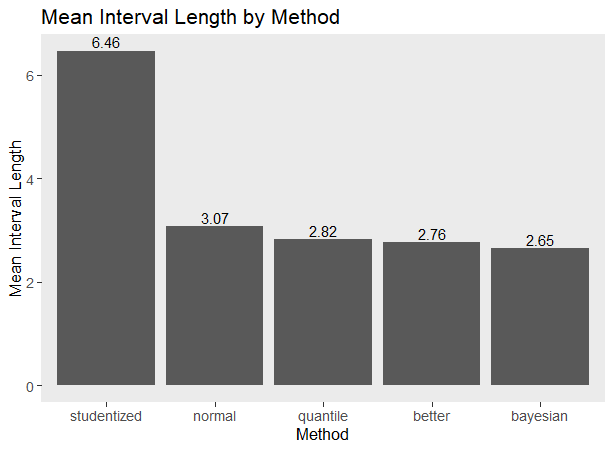}
\label{f23}
\end{figure}

By construction, the median is not affected by those extreme values. For this reason, Figure \ref{f23} depicts the average interval length for each method. Again, the Bayesian method is the better method by this criterion. Furthermore, the Studentized had more than twice the average length of the other methods (Bayesian, Better, Quantile and Normal).

\begin{figure}[!htb]
\caption{Performance indicator by method, using mean rather than median interval length}
\centering
\includegraphics[width=0.6\textwidth]{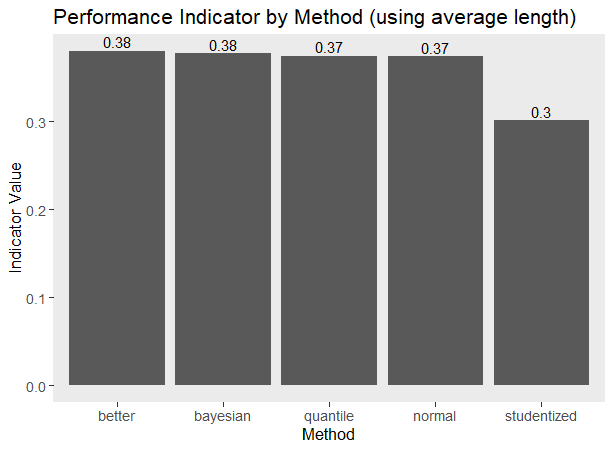}
\label{f24}
\end{figure}

Finally, Figure \ref{f24} shows the behavior of the indicator using the average length instead of the median length. Using this new criterion, the Bayesian and the Better methods produced better results. On the other hand, the Studentized method had a great disadvantage.


\section{Concluding Remarks}

The main goal of this paper was to compare five methods for producing confidence intervals. Four of these methods (Normal, Studentized, Quantile and Better) are directly based on the bootstrap theory developed by \citet{Efron1979} and can be extensively found in the literature. On the other hand, the Bayesian method studied here was constructed inspired by the Bayesian bootstrap method, presented by \citet{Rubin1981}. Although this author did not address the issue of marking confidence intervals, it is possible to do this by relying on the Bayesian notion of credible interval. Thus, we analyzed five methods of generating confidence intervals, one of them constructed based on the Bayesian theory.

To compare the methods, we simulated different scenarios through changes in the sample size, the distribution of the variable and also a autocorrelation induced by a copula model. The methods were applied with different confidence levels and to estimate two different parameters (mean and standard deviation).

The Studentized method produced the best coverage rate, with the simulations showing very similar results to the other methods. This advantage was true even  when individually considering each level of the variables used in the simulation, which is a notable result. However, this advantage became weaker with the increased sample size. For the estimation of the mean, the Normal method generated a very similar coverage rate as the Studentized method. In general, the Bayesian method had the lowest coverage rate, although with little disadvantage compared with the other three methods (Normal, Better and Quantile).

When considering the median length of the intervals, the results were very different: the Bayesian method had the best median interval length, although the difference was not so big in comparison with the other three methods (Normal, Better and Quantile). The main discrepancy was produced by the Studentized method, with clearly larger intervals. Again, the discrepancy between the methods was bigger for small samples and became smaller with the increase of the sample size.

We also proposed an indicator that aims to compare the intervals according to a criterion that considers both the interval length and the coverage rate. All the methods produced very similar results. However, the indicator initially used the median interval length. When we formulated a different version, using the average length, the Bayesian and Better methods performed the best, while the Quantile and Normal methods had small disadvantages; and the Studentized had a big disadvantage. This occurred because the Studentized method produces intervals with extreme lengths with high probability.

In addition to the results related to the comparison between the methods, the paper also has interesting findings on how the confidence intervals are affected by some parameters. First, with three criteria (coverage rate, indicator and presence of outliers), the best results were obtained by the Uniform distribution and the worst results were for the Student-t distribution. It should be noted that the latter distribution has heavy tails because it attributes large probabilities to extreme values. On the other hand, due to its limited support, the Uniform distribution attributes zero probability to extreme values. This is a good indication that heavy tails are bad for the estimation. Furthermore, this analysis also showed that positive autocorrelation can decrease the coverage rate without changing the median interval length, meaning that positive autocorrelation is bad when it is erroneously used in a method that assumes independence. The simulations also showed that the intervals for the standard deviation had lower coverage rate and much larger intervals than those considering  the mean.

In general, this paper shows, especially for moderate sample sizes, that the Studentized method is preferable when the goal is to maximize the coverage rate, while in cases where the interval length is more important, the Studentized method should be avoided, in favor of the Bayesian method. Finally, all the methods had approximately the same performance when considering the coverage rate and median interval length, but when considering the coverage rate and average interval length (more sensitive to the effect of extremely large intervals), the Better and Bayesian methods were the best, and the Studentized method was the worst, so it should be avoided.

\section{Acknowledgments}

Vinícius Litvinoff Justus and Vítor Batista Rodrigues  acknowledge the support provided by CNPq via fellowships from the Master's Program in Statistics of the Universidade Estadual de Campinas, Brazil.

\end{document}